\newcommand{\avu}{\overline{u}} 

\documentclass[pre,twocolumn,showpacs,a4paper,byrevtex]{revtex4}
\usepackage{graphics}

\begin{document}

\title{
  Progressive motion of an ac-driven kink in an annular damped system
}

\author{Edward Goldobin}
\homepage{http://www.geocities.com/e_goldobin}
\email{e_gold@mail.ru}
\affiliation{
  Institut f\"{u}r Schicht- und Ionentechnik (ISI),
  Forschungszentrum J\"ulich GmbH (FZJ), D-52425, J\"ulich, Germany
}

\author{Boris~A.~Malomed}
\email{malomed@eng.tau.ac.il}
\affiliation{
  Department of Interdisciplinary Studies, Faculty of Engineering,
  Tel Aviv University, Tel Aviv 69978, Israel
}

\author{Alexey~V.~Ustinov}
\affiliation{
  Physikalisches Institut, Universit\"{a}t Erlangen-N\"{u}rnberg,
  D-91054, Erlangen, Germany
}

\date{\today}

\begin{abstract}
  A novel dynamical effect is presented: systematic drift of a topological soliton in ac-driven weakly damped systems with periodic boundary conditions. The effect is demonstrated in detail for a long annular Josephson junction. Unlike earlier considered cases of the ac-driven motion of fluxons (kinks), in the present case the long junction is {\em spatially uniform}. Numerical simulations reveal that progressive motion of the fluxon commences if the amplitude of the ac drive exceeds a threshold value. The direction of the motion is randomly selected by initial conditions, and a strong hysteresis is observed. An analytical approach to the problem is based on consideration of the interaction between plasma waves emitted by the fluxon under the action of the ac drive and the fluxon itself, after the waves complete round trip in the annular junction. The analysis predicts instability of the zero-average-velocity state of the fluxon interacting with its own radiation tails, provided that the drive's amplitude exceeds an explicitly found threshold. The result is valid if the phase shift $\varphi$ of the radiation wave, gained after the round trip, is such that $\sin \varphi <0$, the threshold amplitude strongly depending on $\varphi$. A very similar dependence is found in the simulations, testifying to the relevance of the analytical consideration.
\end{abstract}


\maketitle

\section{Introduction}

\label{Sec:intro}

It is well known that progressive motion of topological solitons ({\it kinks}
) in lossy media can be supported by a dc (constant) field which couples to
the soliton's topological charge \cite{review}. In this case, the kink is,
effectively, a quasiparticle. The same quasiparticle approximation strongly
suggests that ac (time-periodic with zero mean value) field(s) applied to
the kink in an infinitely long homogeneous system can only give rise to its
oscillatory motion.

A more sophisticated possibility is to drive a kink by a pure ac force in a
lossy system subjected to a periodic spatial modulation. In particular,
spatial periodicity is an inherent feature of discrete systems (dynamical
lattices). As it was first predicted in an analytical form for damped
lattices of the Frenkel-Kontorova (FK) and Toda types in Refs. \cite{Bonilla}
and \cite{Toda}, respectively, and for a periodically modulated lossy
continuum system of the sine-Gordon (sG) type in Ref. \cite{Bob}, resonances
(of different orders) between the time-periodic driving force and periodic
passage of a moving kink through the spatial inhomogeneities make it
possible to support progressive motion of the kink without any dc field.
This mechanism selects absolute values of the mean velocity of the ac-driven
kink which provide for the resonance of the order $m:n$,
\begin{equation}
u=\frac{m\omega L}{2\pi n},  \label{resonance}
\end{equation}
where $\omega $ and $L$ are the frequency of the ac drive and the period of
the spatial modulation. The sign of the velocity is determined by a random
initial push applied to the kink. While the velocity selected by the locking
condition (\ref{resonance}) does not depend on the amplitude of the ac
drive, the driven progressive motion cannot take place unless the drive's
amplitude $\epsilon $ exceeds a certain minimum (threshold) value $\epsilon
_{{\rm thr}}$, which is usually proportional to the friction coefficient
\cite{Bonilla,Toda,Bob}.

The effect was generalized to include a case when a combination of ac and dc
fields is applied \cite{Mario}. In this case, a {\it constant-velocity step}
is predicted, in the form of an ac-frequency-locked value (\ref{resonance})
of the mean velocity of the driven kink in a finite interval of values of
the dc component of the drive.

The possibility of stable ac-driven motion of the soliton in the lossy Toda
lattice has been later demonstrated in simulations \cite{Jarmo} and in a
laboratory experiment with an electrical transmission line described by this
model \cite{Tom}. Search for such regimes in FK models turned out to be more
difficult, but simulations have finally demonstrated that the soliton in the
form of a dislocation may be driven with a nonzero mean velocity by the ac
force \cite{Giovanni}.

In terms of physical applications, the most appropriate medium where these
effects may be realized is provided by long Josephson junctions (LJJs) \cite
{Bob,Mario}, which support topological solitons in the form of trapped
fluxons (magnetic-flux quanta \cite{Barone}). A convenient way to implement
the periodic spatial modulation in LJJ is to apply constant (dc) magnetic
field to an annular (and sufficiently long) junction, with the field's
vector lying in the plane of the annular junction. In this case, the full
length of the ring-shaped junction plays the role of the modulation period
$L$ in Eq.~(\ref{resonance}). Note that the use of the LJJ of the annular
shape is very natural also because it preserves the number of fluxons
trapped in it. The necessary drive can be readily provided by ac electric
bias current distributed along the junction.

As it was first proposed in Ref. \cite{Denmark}, this configuration gives
rise to a harmonic spatial modulation in the corresponding sG model, the
modulation amplitude being proportional to the external magnetic field (the
perturbation generated by the magnetic field is formally tantamount to an
additional spatially modulated dc drive with zero average). The resonant
relation (\ref{resonance}) applies to this case, the period $L$ being the
circumference of the annular junction, as it was mentioned above.
Experimental observation of the ac-driven motion of a fluxon in the annular
LJJ in the presence of the constant magnetic field was reported in Ref. \cite
{Lyosha}, in the form of nonzero dc voltage across the junction, induced by
ac bias current applied to it.

A related issue is progressive motion of the fluxon under the action of a
pure ac drive, or even a random drive, in the so-called {\it ratchet}
(asymmetric) effective potential \cite{Goldobin:RatchetT}. However, a
principal difference is that, in the case of the ratchet, there is an {\it a
priori} selected preferred direction of motion. Experimental realization of
a ratchet for fluxons in LJJs was very recently reported in Ref.~\cite
{Carapella:RatchetE}.

Thus, ac-driven progressive motion of a topological soliton in an
indefinitely long lossy system can only be possible if it is assisted by the
spatial periodic modulation. The objective of this work is to show, both
numerically and analytically, that, quite surprisingly, the ordinary
spatially uniform ac drive may support progressive, on average, motion of a
fluxon in either direction in the perfectly {\em uniform} annular junction
with non-zero losses, provided that the amplitude of the drive exceeds a
certain threshold value. The effect is clearly stipulated by the finite size
of the ring-like junction, being possible because a fluxon perturbed by the
driving field may emit small-amplitude (quasi-linear) waves (radiation,
alias ``plasmons'' or ``phonons'') propagating along the junction and
eventually hitting the same fluxon.

We note that a somewhat similar effect was observed in numerical simulations
reported in recent preprints \cite{Salerno,Dresden}, which, however, was
related to the above-mentioned ratchet mechanism supporting unidirectional
motion of ac-driven fluxons, and the ac drive was always a two-frequency
one. It was stated that the progressive motion is possible due to the
simultaneous action of driving forces oscillating at frequencies $\omega$
and $2\omega$ in the absence of static spatial potential. We stress that in
our work the ac drive is always represented by a single-frequency harmonic.

The rest of the paper is organized as follows. Numerical results which show
the ac-driven progressive motion of the fluxon are displayed in section \ref
{Sec:SimRes}. A theoretical explanation, that reflects our understanding of
the effect, which is based on interaction of the fluxon with its own
radiation ``tails'', is set forth in section \ref{Sec:Theory}. Comparison of
the analytical and numerical results is presented in section \ref{Sec:Disc},
and section \ref{Sec:Conclusion} concludes the work.

\section{Numerical results}

\label{Sec:SimRes}

\subsection{The model and computational procedure}

The ac-driven weakly damped annular Josephson junction is described by the
well-known perturbed sG equation for the Josephson phase $\phi (x,t)$ \cite
{Barone},
\begin{equation}
\phi _{xx}-\phi _{tt}-\sin \phi =\alpha \phi _{t}-\gamma _{{\rm ac}}\sin
(\omega t),  \label{Eq:sG}
\end{equation}
where the subscripts stand for the partial derivatives, the coordinate $x$,
running along the junction, and the time $t$ are normalized so that the
Josephson penetration length and plasma frequency are both equal to $1$,
$\alpha $ is the damping coefficient, and $\gamma _{{\rm ac}}$ is the
amplitude of ac drive. In the infinite system without perturbations ($\alpha
=\gamma _{{\rm ac}}=0$), the fluxon is described by the kink solution to the
sG equation,
\begin{equation}
\phi _{{\rm fl}}=4\arctan \left[ \exp \left( \sigma \frac{x-\xi (t)}{\sqrt{
1-u^{2}}}\right) \right] ,  \label{Eq:fluxon}
\end{equation}
where $u$ is the velocity of the fluxon, $\sigma =\pm 1$ is its polarity,
and $\xi $ is the instantaneous coordinate of its center, which is $\xi
(t)=\xi _{0}+ut$ in the case of free motion of the fluxon. If a single
fluxon is trapped in the annular junction, Eq.~(\ref{Eq:sG}) is supplemented
by the boundary conditions (b.c.)
\begin{eqnarray}
\phi (L,t) &=&\phi (0,t)+2\pi ,  \label{Eq:BC1} \\
\phi _{x}(L,t) &=&\phi _{x}(0,t),  \label{Eq:BC2}
\end{eqnarray}
where $L$ is the circumference of the ring. \newline

Equation~(\ref{Eq:sG}) with b.c. (\ref{Eq:BC1}) and (\ref{Eq:BC2}) was
simulated by means of the ``StkJJ'' software package, described in detail
elsewhere \cite{StkJJ}. The results have been checked by comparing them with
simulations carried out with the ``Soliton'' package \cite{backbend-99}, and
good agreement was found in all the cases considered.

The simulations were performed as follows. We started with an initial state
corresponding to a fluxon trapped in the system at the zero amplitude of ac
drive. Then the amplitude was increased by small steps $\Delta \gamma _{{\rm
ac}}$, each time calculating the average dc voltage $V_{{\rm dc}}$ across
the junction,
\begin{eqnarray}
  V_{{\rm dc}}&\equiv&\frac{1}{2\pi TL}\int_0^T dt\int_0^L dx\,\phi_t(x,t)
  \nonumber\\
  &\equiv&\frac{1}{2\pi TL}\int_0^L dx\,\left[\phi(x,T)-\phi(x,0)\right]
  ,  \label{Eq:v_av}
\end{eqnarray}
where $T$ is a sufficiently large time-averaging interval [in physical
units, the average voltage is given by the expression (\ref{Eq:v_av})
multiplied by the magnetic flux quantum $\Phi _{0}$]. The value of $T$ was
taken as multiple of the ac-drive's period $2\pi /\omega $.

The dependence $V_{{\rm dc}}(\gamma _{{\rm ac}})$ characterizes the average
velocity of the progressive motion of the fluxon in the annular junction.
Indeed, substituting a solution for the moving fluxon in the form $\phi
(x,t)=$ $\phi _{{\rm fl}}(x-\xi (t))$, see Eq.~(\ref{Eq:fluxon}), into the
first integral expression in Eq.~(\ref{Eq:v_av}), we obtain
\begin{eqnarray}
  V_{\rm dc}&\equiv&\frac{1}{2\pi TL}\int_{0}^{T}dt\frac{d\xi }{dt}
  \int_{0}^{L}dx\,\phi _{{\rm fl}}^{\prime }\left( x-\xi (t)\right)
  \nonumber\\
  &=&\frac{1}{2\pi TL}\int_{0}^{T}dt\frac{d\xi }{dt}\left[ \phi _{{\rm fl}}\left(L,t\right) -\phi _{{\rm fl}}\left( 0,t\right) \right]
  \nonumber\\
  &=&\frac{1}{TL}\int_{0}^{T}dt\frac{d\xi }{dt}
  ,  \label{Vdc}
\end{eqnarray}
where b.c.~(\ref{Eq:BC2}) was used. A natural definition of the fluxon's
average velocity being
\[
\avu=\frac{1}{T}\int_{0}^{T}dt\frac{d\xi }{dt}\,,
\]
Eq.~(\ref{Vdc}) yields an expression for the average velocity in terms of
the average dc voltage,
\begin{equation}
\avu=LV_{{\rm dc}}\,.  \label{Eq:V(u)}
\end{equation}
The results of the simulations are displayed below in the form $\avu=
\avu(\gamma _{{\rm ac}})$, using numerically found dependencies $V_{
{\rm dc}}(\gamma _{{\rm ac}})$ and the relation (\ref{Eq:V(u)}). This way of
the presentation of results is the most appropriate one if the objective is
to look for dynamical regimes with a nonzero average velocity of the
ac-driven kink.

For some values of $\gamma _{{\rm ac}}$ the motion of fluxon is apparently
chaotic (see below). In such cases, the average velocity calculated with
Eq.~{\ref{Vdc}} will not converge as $T\rightarrow \infty$. To avoid this
problem, we used two different averaging procedures:

\begin{itemize}
\item  (i) The averaging was made simply over one period $2\pi /\omega $ of
the ac drive, and the amplitude $\gamma _{{\rm ac}}$ was changed in
extremely small steps, $\delta \gamma _{{\rm ac}}=0.0001$, while, otherwise,
$\delta \gamma _{{\rm ac}}=0.01$ is used. This, in a sense, means that
instead of one point on the $\avu(\gamma _{a}c)$ dependence we get
100 closely situated points. If the dynamics is chaotic those 100 points
will have large spread in $\avu$. On the other hand, all the points
will give almost equal values of $\avu$ if the dynamics is regular
and averaging converges. Although some artifacts may appear around
bifurcation points, this technique provides very good insight into dynamical
states of the system.

\item  (ii) The averaging was performed iteratively for gradually increasing
value of $T$ until the convergence of $\avu$ with the accuracy $0.001
$ was reached, but not longer than for $T=1000$. In this case, for the
non-chaotic states the averaging gives the true value of equilibrium average
velocity $\avu$, while for chaotic states averaging was interrupted
at $T=1000$, and the corresponding non-converged value of $\avu$ was
included into the plot.
\end{itemize}

Results produced by means of both methods will be displayed below.

\subsection{The ac-driven progressive motion}

\begin{figure}[tb]
  \centering\includegraphics{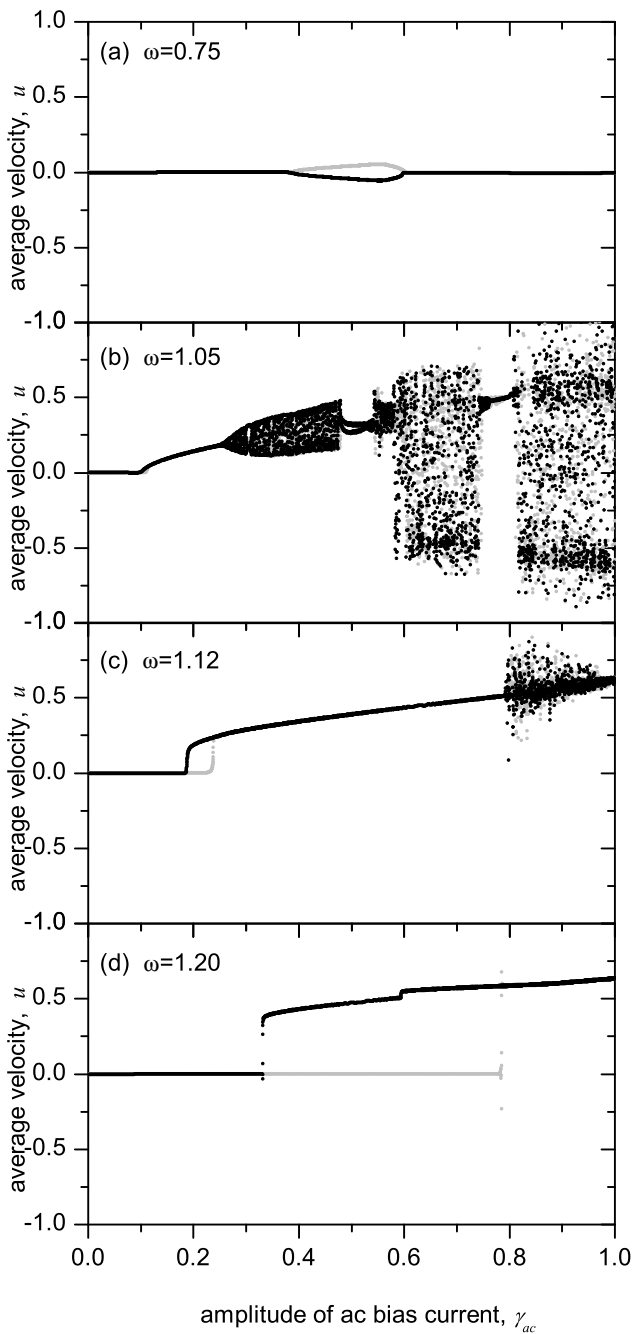}
  \caption{
    The dependences $\avu(\protect\gamma_{{\rm ac}})$ obtained by sweeping $\protect\gamma_{{\rm ac}}$ up (black points) and down (gray points) at four different frequencies of the ac drive: $\protect\omega=0.75$ (a), $\protect\omega=1.05$ (b), $\protect\omega=1.12$ (c), and $\protect \omega=1.20$ (d). The length of the annular junction is $L=10$, and the damping coefficient is $\protect\alpha=0.1$.
  }
  \label{Fig:u(gamma)}
\end{figure}

First, we present dependencies $\avu(\gamma _{{\rm ac}})$ generated
by the averaging method (i). Four characteristic examples of the
dependencies, obtained by increasing (gray points) and decreasing (black
points) the drive's amplitude $\gamma _{{\rm ac}}$ for the annular LJJ of
the length $L=10$, with the dissipation constant $\alpha =0.1$ and four
different values of the driving frequency $\omega $, are shown in Fig.~\ref
{Fig:u(gamma)}. In Fig.~\ref{Fig:u(gamma)}(a), corresponding to $\omega =0.75
$, the average velocity of the fluxon remains zero in the region $\gamma _{
{\rm ac}}<0.38$. However, in the interval $0.38<\gamma _{{\rm ac}}<0.6$ the
average velocity is {\em different} from zero, which means that the fluxon
moves, on average, in one direction. As one sees in Fig.~\ref{Fig:u(gamma)}
(a), the signs of the average velocity are different for the branches of the
characteristic $\avu(\gamma _{{\rm ac}})$ corresponding to
``sweeping up'' and ``sweeping down'', i.e., increasing and decreasing
$\gamma _{{\rm ac}}$. Due to the symmetry of the system, these signs are
actually picked up randomly, depending on the initial conditions and
peculiarities of the simulations, such as the way $\gamma _{{\rm ac}}$ was
varied.

At the frequency $\omega =1.05$, corresponding to Fig.~\ref{Fig:u(gamma)}(b),
the fluxon is set into motion in the region $\gamma _{{\rm ac}}>0.1$.
The average velocity $\avu$ increases with $\gamma _{{\rm ac}}$, but
at $\gamma _{{\rm ac}}\approx 0.26$ a transition to a chaotic-like behavior
takes place. In the latter case, the averaging over a long time interval
does not produce any definite value of the mean velocity. Closer inspection
of this region shows that we are dealing not with true chaos, but rather
with a deterministic quasi-periodic regime with a nonzero average velocity.
In particular, the fundamental frequency of the corresponding time series
is, effectively, incommensurate with the ac driving frequency in this case,
that is why the average velocity cannot be effectively calculated using the
algorithm described above.

Continuing the consideration of Fig.~\ref{Fig:u(gamma)}(b), we notice a
split branch in the interval $0.48<\gamma _{{\rm ac}}<0.54$, where the
average velocity is $\avu\approx 0.3$. Accurate inspection shows
that the averaging during one period yields a point belonging to one branch,
while averaging over the next period produces a point situated on the other
branch, and so on. This means that the fundamental frequency of the
oscillatory component of the fluxon's motion is {\em half} the driving
frequency. While generation of higher harmonics is common to all nonlinear
systems, the appearance of sub-harmonics is a more specific effect, implying
the existence of a certain form of parametric instability. For larger values
of $\gamma _{{\rm ac}}$, the system enters the region of chaos, which is
shortly interrupted by a deterministic branch with a nonzero average
velocity in the interval $0.75<\gamma _{{\rm ac}}<0.80$.

For driving frequencies that are $\sim 10\%-20\%$ higher than the Josephson
plasma frequency (which is $1$ in the notation adopted above), one can
observe the fluxon's motion with a nonzero average velocity in a broad range
of $\gamma _{{\rm ac}}$, see Figs.~\ref{Fig:u(gamma)}(c) and (d). It is
noteworthy that the average velocity may attain relatively large values,
which become comparable to the Swihart velocity of the junction ($\equiv 1$
in our notation). For instance, in Fig.~\ref{Fig:u(gamma)}(d) $\avu
\approx 0.5$ at $\gamma _{{\rm ac}}=0.5$, and reaches the value $\approx 0.7$
for larger values of $\gamma _{{\rm ac}}$.

It is also interesting to note that the dependence $u(\gamma _{{\rm ac}})$
shown in Fig.~\ref{Fig:u(gamma)}(d) features an additional step at $\gamma _{
{\rm ac}}\approx 0.6$, which suggests that there are two different
mechanisms which generate the nonzero average velocity. The additional step
may correspond to switching of the system between these two different
mechanisms.

\begin{figure*}[tb]
  \centering\includegraphics{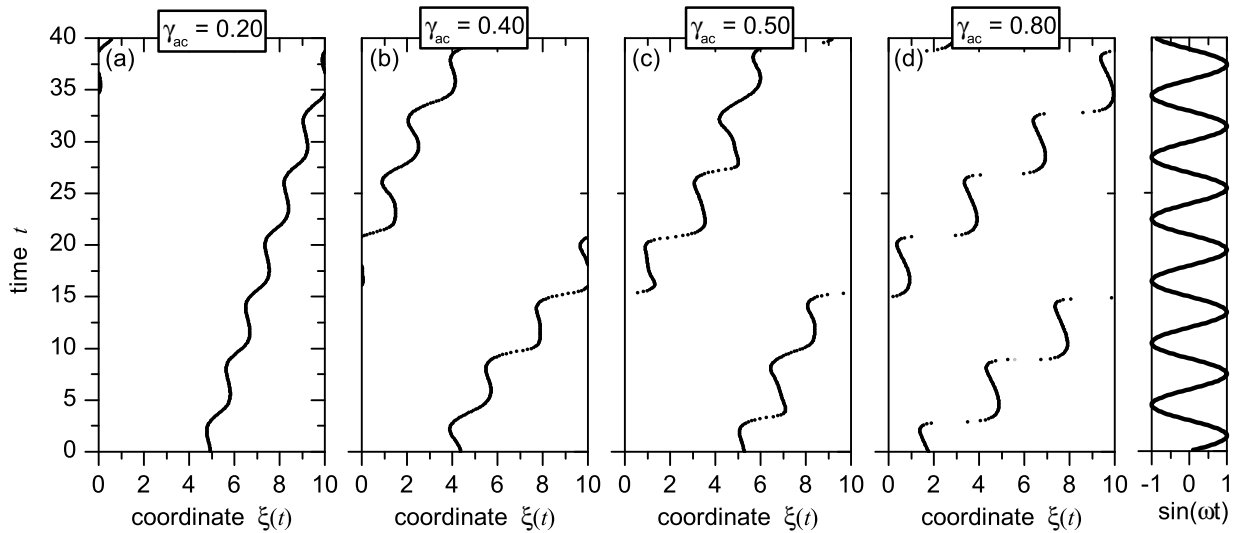}
  \caption{
    The coordinate of the fluxon's center as a function of time for the ac-drive's frequency $\protect\omega=1.05$ [which corresponds to the case shown in Fig.~\ref{Fig:u(gamma)}(b)] and different values of the drive's amplitude $\protect\gamma_{{\rm ac}}$: $0.2$ (a), $0.4$ (b), $0.5$ (c), and $0.8$ (d). The rightmost plot shows the driving ac signal $\sin( \protect\omega t)$ vs. time.
  }
  \label{Fig:x(t)}
\end{figure*}

To visualize the fluxon motion, in Fig.~\ref{Fig:x(t)} we present the law of
motion, $\xi (t)$, of the fluxon's center for different values of the ac
driving amplitude and fixed driving frequency $\omega =1.05$, which
corresponds to the dependence $\avu(\gamma _{{\rm ac}})$ shown in
Fig.~\ref{Fig:u(gamma)}(b). It is clearly seen in Fig.~\ref{Fig:x(t)}(a)
that, indeed, at $\gamma _{{\rm ac}}=0.2$ a systematic drift of the fluxon
in one direction is superimposed on periodic oscillations. In Fig.~\ref
{Fig:x(t)} (b), which corresponds to the quasi-periodic motion at $\gamma _{
{\rm ac}}=0.4$, we see that the fluxon motion is qualitatively similar to
that displayed in Fig.~\ref{Fig:x(t)}, but it is less regular and
nonperiodic.

The fluxon's law of motion for the case which corresponds to the generation
of the sub-harmonic at $\gamma _{{\rm ac}}=0.5$ is shown in Fig.~\ref
{Fig:x(t)}(c). One can easily see that the period\ of the oscillatory part
of the soliton's motion in this case is indeed twice the period of the ac
drive, as the trajectories look slightly different during alternating (even
and the odd) periods of the ac drive.

Finally, in Fig.~\ref{Fig:x(t)}(d) one can see the trajectories
corresponding to the point $\gamma _{{\rm ac}}=0.8$, which belongs to a
narrow window in Fig.~\ref{Fig:u(gamma)}(b), where a deterministic law of
motion becomes stable again. Qualitatively, this plot is similar to that in
Fig.~\ref{Fig:x(t)}(a), but has a higher average velocity.

\begin{figure}[tb]
  \centering\includegraphics{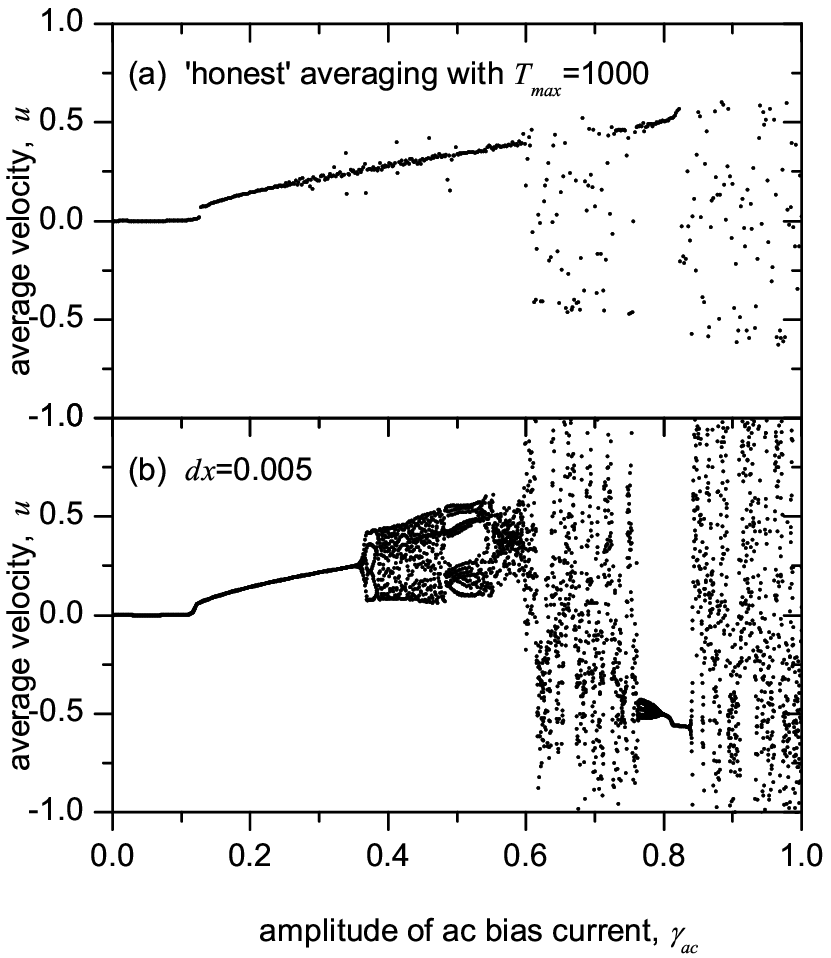}
  \caption{
    The dependence $\avu(\protect\gamma _{ac})$ produced using: (a) the alternative averaging procedure, cf. Fig.~\ref{Fig:u(gamma)}(b); and (b) a smaller spatial step of the finite-difference integration scheme. The scattered points in panel (a) are those at which the averaging did not converge, thus they mark regions of the chaotic motion of the ac-driven fluxon.
  }
  \label{Fig:cmp-diff-av}
\end{figure}

Results produced by the averaging method (ii) are displayed, for $\omega
=1.05$, in Fig.~\ref{Fig:cmp-diff-av}(a). As is seen from the comparison of
this figure with Fig.~\ref{Fig:u(gamma)}(b), details of the pictures
generated by the averaging procedures (i) and (ii) are somewhat different
inside the chaotic regions. Nevertheless, both methods identify the same
regions of the regular motion, and yield the same dependencies $\avu
(\gamma _{{\rm ac}})$ in the regular cases. In fact, the consideration of
the regular motion with nonzero average velocity, rather than of chaotic
regimes, is the main subject of this work.

Our numerical simulations of the perturbed sG equation (\ref{Eq:sG}) used a
finite-difference scheme. Since the ac-driven progressive motion of kinks is
possible in indefinitely long discrete lossy lattices \cite
{Bonilla,Toda,Jarmo,Giovanni}, it is necessary to check that the effects
reported above are not artifacts produced by the discretization of the
model. To this end, simulations were run with the smaller value of the step
size $\Delta x$ of the finite-difference scheme. Comparing the results, we
have concluded that they are almost identical, except for small variations
in the location of the thresholds at which the system switches from one
state to the other. For example, Fig.~\ref{Fig:cmp-diff-av}(b) displays the
dependence $\avu(\gamma _{{\rm ac}})$ obtained for the same case as
in Fig.~\ref{Fig:u(gamma)}(b), but with $\Delta x=0.005$, rather than
$\Delta x=0.01$ used in Fig.~\ref{Fig:u(gamma)}(b). As for the different
signs of the average velocity in some regions of the regular motion in Figs.~
\ref{Fig:u(gamma)}(b) and \ref{Fig:cmp-diff-av}(b), this sign is always
randomly selected by the initial conditions, as it was explained above.

The numerical results presented above clearly demonstrate that progressive
motion of fluxon can indeed be supported, in either direction, by the
uniformly distributed ac drive (bias current) in annular uniform weakly
damped LJJ. An explanation to the effect will be proposed in the next
section.

\section{Analytical consideration}

\label{Sec:Theory}

\subsection{Radiation waves emitted by the ac-driven fluxon}

The analysis developed below is based on a basic property of the system
under consideration: the fluxon oscillating under the action of the ac drive
emits plasma waves in both directions, and, due to the annular shape of the
junction, these waves interact with the same fluxon after completing a round
trip. However, this property does not offer an immediate explanation to the
possibility that the fluxon will be able to systematically drift in one
direction, if pushed initially. Therefore, a detailed analysis is necessary,
which is developed below.

As it is known from earlier works on the perturbation theory (PT) for the sG
equation (see a review \cite{review}), a kink (fluxon) oscillating, without
any systematic progressive motion, under the action of the ac drive emits
radiation, in the lowest approximation, to the left and to the right at two
wavenumbers $\pm k$, where
\begin{equation}
k=\sqrt{\omega ^{2}-1},  \label{k}
\end{equation}
provided that $\omega >1$, i.e., that the driving frequency exceeds the
plasma frequency of the junction. If $\omega <1$, the emission takes place
in higher orders of PT (at higher harmonics).

Here, we concentrate on the case $\omega >1$. The numerical results (both
displayed in Fig.~\ref{Fig:u(gamma)} and those which are not shown here)
suggest that the strongest effect takes place when $\left( \omega
^{2}-1\right) $ is a positive but relatively small parameter. In this case,
the application of PT, which needs the drive's amplitude $\gamma _{{\rm ac}}$
and the damping coefficient $\alpha $ to be small, predicts that the
oscillating fluxon emits two waves, which, far from the fluxon, have the
following form (neglecting, for the time being, the action of the
dissipation):
\begin{equation}
\phi _{\pm }=A_{0}\sin \left( \pm k\,x-\omega t\right) .  \label{wave}
\end{equation}
The amplitude of the two emitted waves is predicted by PT to be \cite{review}
\begin{equation}
A_{0}=\gamma _{{\rm ac}}/k^{2},  \label{amplitude}
\end{equation}
which is correct if
\begin{equation}
k^{2}\gg \gamma _{{\rm ac}}^{2/3}\,\,{\rm and}\,\,k^{2}\gg \alpha .
\label{conditions}
\end{equation}
In fact, the former inequality does not hold in many cases when the
simulation reveal the progressive motion of the ac-driven fluxon, but this
is not a principal limitation: instead of the expression (\ref{amplitude}),
one can use more general ones (6.23) and (6.18) from Ref. \cite{review}. In
such a case, a relation between $A_{0}$ and the drive's amplitude $\gamma _{
{\rm ac}}$ is more complex than that given by Eq.~(\ref{amplitude}), but
there will be no essential change in the analysis presented below, nor
significant changes in the final results.

Attenuation of the emitted waves due to the presence of the loss term in
Eq.~(\ref{Eq:sG}) will play a crucially important role below. To consider
the attenuation, it is convenient to use a reference frame ($x^{\prime
},t^{\prime }$) moving at a constant velocity $\avu$, which, as well
as in the previous section, is the average velocity of the fluxon (if it
moves on average). In the analysis, we will treat $\avu$ as the
smallest parameter, in comparison with $\gamma _{{\rm ac}}$, $\alpha $, and
$\omega ^{2}-1\equiv k^{2}$, the main objective being to show that, at a
certain threshold value $\left( \gamma _{{\rm ac}}\right) _{{\rm thr}}$ of
the ac-drive's amplitude, the state with $\avu=0$ becomes {\em
unstable}, and a symmetric pair of new nontrivial states with finite but
very small average velocities $\pm \,\avu$ appears as a result of an
instability-triggered {\it pitchfork bifurcation} \cite{bifurcation} if
$\gamma _{{\rm ac}}$ slightly exceeds $\gamma _{{\rm th}}$.

In the moving reference frame, the linearized version of Eq.~(\ref{Eq:sG}),
which governs the propagation and attenuation of the radiation waves,
becomes
\begin{equation}
\phi _{x^{\prime }x^{\prime }}-\phi _{t^{\prime }t^{\prime }}-\phi =\frac{
\alpha }{\sqrt{1-\avu^{2}}}\phi _{t^{\prime }}-\frac{\alpha v}{\sqrt{
1-\avu^{2}}}\phi _{x^{\prime }}\,,  \label{linear}
\end{equation}
where $x^{\prime }=(x-{\avu}t)/\sqrt{1-\avu^{2}}$ and
$t^{\prime }=(t-{\avu}x)\sqrt{1-\avu^{2}}$. A stationary shape
of the radiation wave in the moving reference frame is sought for as [cf.
Eq.~(\ref{wave})]
\begin{equation}
\phi _{\pm }=A_{\pm }(x^{\prime })\sin \left( \pm \sqrt{\omega ^{\prime
}{}^{2}-1}\cdot x^{\prime }-\omega ^{\prime }t^{\prime }\right) ,
\label{wave'}
\end{equation}
where a slow dependence $A(x^{\prime })$ is produced by the dissipative
damping of the wave, and $\omega ^{\prime }$ is the driving frequency in the
moving reference frame,
\begin{equation}
\omega ^{\prime }=\frac{\omega }{\sqrt{1-\avu^{2}}}\,.  \label{omega}
\end{equation}
In the first approximation, which assumes that both the loss constant
$\alpha $ and the average velocity $\avu$ are small, the substitution
of Eq.~(\ref{wave'}) into Eq.~(\ref{linear}) gives rise to an equation
\begin{equation}
\frac{dA_{\pm }}{dx^{\prime }}=\mp \frac{\alpha }{2u_{{\rm gr}}^{\prime }}
\left( 1\pm u_{{\rm gr}}^{\prime }\avu\right) A_{\pm },
\label{Eq:dA/dx}
\end{equation}
where the notation for the radiation-wave's group velocity was introduced
\begin{equation}
u_{{\rm gr}}^{\prime }\equiv \frac{d\omega ^{\prime }}{dk^{\prime }}=\sqrt{
\omega ^{\prime }{}^{2}-1}/\omega ^{\prime }\approx k\left[ 1+\avu
^{2}/\left( 2k^{2}\right) \right] .  \label{ugr}
\end{equation}
{\bf \ }To obtain the final expression in Eq.~(\ref{ugr}), it was assumed
that, in accord with what was said above, $\avu^{2}\ll k^{2}\ll 1$
(hence the second term in the square brackets is a small correction).

A solution to Eq.~(\ref{Eq:dA/dx}) is obvious:
\begin{equation}
A_{\pm }(x^{\prime })=A_{0}\exp \left[ -\frac{\alpha }{2u_{{\rm gr}}^{\prime
}\sqrt{1-\avu^{2}}}\left( 1\pm u_{{\rm gr}}^{\prime }\avu
\right) |x^{\prime }|\right] .  \label{A(x)}
\end{equation}
In particular, we will need the value of the emitted-wave's amplitude after
the completion of the round trip, from $x^{\prime }=0$ to $x^{\prime
}=L^{\prime }=L\sqrt{1-\avu^{2}}$, where the Lorentz contraction is
taken into regard. According to Eqs.~(\ref{A(x)}) and (\ref{ugr}), we find
\begin{equation}
A_{\pm }(L^{\prime })=\widetilde{A}_{0}\left( 1\mp \frac{1}{2}\alpha
L\avu\right) ,  \label{final}
\end{equation}
where
\begin{equation}
\widetilde{A}_{0}\equiv A_{0}\exp \left( -\frac{\alpha L}{2u_{{\rm gr}
}^{\prime }}\right) \approx A_{0}\exp \left( -\frac{\alpha L}{2k}\right)
\cdot \left( 1+\frac{\alpha L\avu^{2}}{4k^{3}}\right) \,
\label{tilde}
\end{equation}
(hereafter, it is implied that $k\equiv \sqrt{\omega ^{2}-1}$ is taken as a
positive square root). As it was said above, the fundamental assumption in
the analysis is that the velocity $\avu$ is very small.
Nevertheless, small corrections $\sim \avu$  and $\sim \avu
^{2}$, which are retained in Eqs. (\ref{final}) and (\ref{tilde}),
respectively, will play an important role below.

\subsection{ Dragging the fluxon by the radiation waves in the circular
system}

As we are dealing with the circular system, each emitted wave, after having
completed its round trip, strikes the fluxon and exerts some dragging force
acting on it. Dragging a kink by a radiation wave passing through it was
analyzed in detail in Ref.~ \cite{Japan}. It was found that a nonzero
dragging force appears at the second order in the dragging-wave's amplitude,
provided that $\alpha \neq 0$. An expression for the dragging forces induced
by the two waves $\phi _{\pm }$ can be shown to take the following form:
\begin{equation}
F_{{\rm drag}}^{\pm }=\pm \left( \pi /4\right) ^{2}\alpha kA_{\pm }^{2}
\label{Eq:F-drag}
\end{equation}
(this expression assumes that the kink's mass is normalized to be $1$).
However, taking the expressions (\ref{final}) for the amplitudes $A_{\pm }$
and substituting them into the expression (\ref{Eq:F-drag}), one can readily
see that the resultant force, produced by the asymmetry between the two
waves, would be not accelerating, but {\em braking} the fluxon's motion.

A key point for the explanation of the possibility of the ac-driven
progressive motion is to notice that the waves (\ref{wave'}) act on the
fluxon in combination with the direct ac drive. To take all the forces into
regard, a known perturbative technique may be applied \cite{review}: the net
wave field is sought for in the form
\begin{equation}
\phi (x,t)=\psi (x,t)+\phi _{0}(t)+\phi _{+}(x,t)+\phi _{-}(x,t),
\label{combination}
\end{equation}
where $\psi (x,t)$ is the field of the fluxon, $\phi _{\pm }$ are the
emitted waves (\ref{wave}), and
\begin{equation}
\phi _{0}(t)\approx -\left( \gamma _{{\rm ac}}/k^{2}\right) \sin (\omega t)
\label{phi0}
\end{equation}
is a term representing uniform background oscillations generated by the ac
drive, as it follows from a straightforward solution to the linearized
equation (\ref{Eq:sG}) (recall that $k\equiv \sqrt{\omega ^{2}-1}$). Then,
an effective equation for the fluxon field $\psi $ is obtained by the
substitution of the combination (\ref{combination}) into the underlying
perturbed sG equation (\ref{Eq:sG}), expansion of $\sin \phi $, and taking
into regard the expression (\ref{phi0}) for the background component of the
field. The final result (written in the laboratory reference frame) is
\begin{eqnarray}
  &&\psi _{tt}-\psi _{xx}+\sin \psi +\alpha \psi _{t} =
  -\left(1-\cos\psi\right)\cdot  \{
    \nonumber\\
  &&\widetilde{A}_{0}\left[
     \sin\left(kx-\omega t+kL\right)+
     \sin\left(-kx-\omega t+kL\right)
  \right]
  \nonumber \\
  &&-\frac{1}{2}\alpha L\widetilde{A}_{0}\avu\left[
    \sin \left( kx-\omega t+kL\right) -
    \sin \left(-kx-\omega t+kL\right)
  \right]
  \nonumber\\
  &&\left(\gamma_{\rm ac}/k^2\right)\sin\left(\omega t\right)
  \}
  , \label{psi}
\end{eqnarray}
where the amplitude $\widetilde{A}_{0}$ is defined by Eq.~(\ref{tilde}), and
the corrections $\mp (1/2)\alpha L\avu$ from Eq.~(\ref{final}) are
taken into account. The phase shift $kL$ appearing on the right-hand side of
Eq.~(\ref{psi}) is accumulated by the wave after the completion of the round
trip along the ring. Note that the renormalization of the driving amplitude
in the first term on the right-hand side of Eq.~(\ref{psi}) is quite
essential in the present case, when $k^{2}\equiv \omega ^{2}-1$ is small.

It is straightforward to derive in the lowest (adiabatic) approximation \cite
{review} an equation of motion for the coordinate $\xi $ of the fluxon
driven by the terms on the right-hand side of Eq.~(\ref{psi}). In the
``non-relativistic' approximation (for small velocities), it takes the form
\begin{eqnarray}
  &&\frac{d^{2}\xi }{dt^{2}}+\alpha \frac{d\xi }{dt} =\frac{\sigma \pi }{4}
    \cdot \left( \gamma _{{\rm ac}}/k^{2}\right) \sin \left( \omega t\right)
    \nonumber \\
  &&+\widetilde{A}_{0}\left[ \sin \left( \omega t-kL-k\xi \right)
    +\sin \left(\omega t-kL+k\xi \right) \right]
    \nonumber \\
  &&-\frac{1}{2}\alpha L\widetilde{A}_{0}\avu
  \left[ \sin \left( \omega t-kL-k\xi \right) -\sin \left( \omega t-kL+k\xi \right) \right]
  ,\label{driving}
\end{eqnarray}
[recall that $\sigma =\pm 1$ is the fluxon's polarity, see Eq.~(\ref
{Eq:fluxon})].

The right-hand side (r.h.s.) of Eq.~(\ref{driving}) may be expanded for
small $\xi $. Then one concludes that the first term and the following pair
of terms on r.h.s. are essentially similar to each other (in the lowest
approximation, they do not contain $\xi $), but, with regard to Eqs.~(\ref
{amplitude}) and (\ref{tilde}), the amplitude in front of the latter pair of
terms, $\widetilde{A}_{0}$, differs from the amplitude in front of the first
term by the exponential factor $\exp \left[ -\alpha L/\left( 2k\right)
\right] $, that we assume to be small enough. Therefore, this pair may be
neglected, and, keeping a term linear in $\xi $ which is produced by the
expansion of the last pair of terms on r.h.s. in Eq.~(\ref{driving}), we
arrive at a simplified equation of motion for the driven fluxon,
\begin{equation}
  \frac{d^{2}\xi }{dt^{2}}+\alpha \frac{d\xi }{dt} =
  \frac{\sigma \pi }{4}\left[ \frac{\gamma _{{\rm ac}}}{k^{2}}\sin
  \left( \omega t\right) +\alpha L\widetilde{A}_{0}k\avu\cos \left(
  \omega t-kL\right)\xi \right]
  \label{simplified}
\end{equation}

We seek for a solution to Eq.~(\ref{simplified}) in a natural form (similar
to that employed in Ref. \cite{Japan}),
\begin{equation}
\xi (t)=-\frac{\sigma \pi \gamma _{{\rm ac}}}{4k^{2}}\sin \left( \omega
t\right) +\xi _{0}(t),  \label{Japan}
\end{equation}
where the first term is a response to the ac driving force on r.h.s. of Eq.~
(\ref{simplified}) (we neglect a small correction to it from the friction
force on the left-hand side of the equation, and set $1/\omega ^{2}\approx 1$,
which is true in the case under consideration), and the term $\xi _{0}(t)$
takes into regard a possibility of a slow systematic motion ({\it drift}) of
the fluxon.

In the next approximation, we replace the multiplier $\xi (t)$ in the second
term on r.h.s. of Eq.~(\ref{simplified}) by the first term from the
expression (\ref{Japan}). In order to single out terms contributing to the
slow drift, we perform averaging in the rapid oscillations. Thus we arrive
at an effective evolution equation for the slow variable $\xi _{0}$, in
which $d\xi _{0}/dt$ may be identified by the average velocity $\avu$
of the systematic motion of the fluxon. Finally, the equation for the slow
motion can be cast into the form of a first-order evolution equation for the
average velocity,
\begin{equation}
\frac{d\avu}{dt}=-\alpha \avu-\frac{\pi ^{2}L\widetilde{A}
_{0}\,\gamma ^{{\rm (ac)}}}{32k}\,(\alpha \avu)\sin \left( kL\right)
\,.  \label{slow}
\end{equation}
The amplitude $\widetilde{A}_{0}$ in Eq.~(\ref{slow}) can be replaced by the
expressions (\ref{amplitude}) and (\ref{tilde}), which casts the equation
into a form
\begin{eqnarray}
\frac{d\avu}{dt} &=&\left[ \frac{\left( \gamma _{{\rm ac}}\right)
^{2}}{\gamma _{{\rm thr}}^{2}}-1\right] \alpha \avu  \nonumber \\
&&-\frac{1}{2}\left( \frac{\pi L\alpha \gamma _{{\rm ac}}}{8k^{3}}\right)
^{2}\exp \left( -\frac{\alpha L}{2k}\right) \sin \left( kL\right) \cdot
\avu^{3},  \label{motion}
\end{eqnarray}
where the threshold value of the drive's amplitude is defined as
\begin{equation}
\gamma _{{\rm thr}}^{2}\equiv -\frac{\left( 32/\pi ^{2}\right) k^{3}}{L\sin
\left( kL\right) }\exp \left( \frac{\alpha L}{2k}\right) .  \label{threshold}
\end{equation}

The equation of motion (\ref{motion}) has a trivial stationary solution
(trivial {\it fixed point}, FP) $\avu=0$, which, obviously,
corresponds to no systematic motion of the fluxon. A crucially important
issue is the stability of this FP. The linearization of Eq.~(\ref{motion})
for $\avu\rightarrow 0$ immediately demonstrates that the trivial FP
is {\em unstable} if the drive's amplitude exceeds the threshold value given
by Eq.~(\ref{threshold}). In this case, a transition to a nontrivial regime
with a {\em nonzero velocity} of the progressive motion (in other words, a
bifurcation) must take place.

An essential feature of the expression (\ref{threshold}) is its dependence
on the phase $kL$ gained by the emitted wave as a result of the round trip.
In fact, the ac-driven motion is predicted to exist only if $\gamma _{{\rm
thr}}^{2}>0$, or, as it follows from Eq.~(\ref{threshold}),
\begin{equation}
\sin \left( kL\right) <0.  \label{condition}
\end{equation}
In the opposite case, the effect is absent, or, possibly, the threshold
amplitude is very large (then the perturbative approach is irrelevant).

If $\gamma _{{\rm ac}}^{2}>\gamma _{{\rm thr}}^{2}>0$, the existence of two
mutually symmetric nontrivial FPs, with finite velocities $\pm \,\avu
$, may be expected. One can try to find these velocities close to the
bifurcation point, i.e., for a case $0<\gamma _{{\rm ac}}^{2}-\gamma _{{\rm
thr}}^{2}\ll \gamma _{{\rm thr}}^{2}$, taking into account the cubic
corrections in Eq.~(\ref{motion}). In fact, in the particular approximation
in which the cubic term, which is present in Eq.~(\ref{motion}), was
derived, a formal result will be $\avu^{2}<0$, which, actually,
implies that the bifurcation is {\it subcritical}, i.e., it takes place by a
finite jump; in the opposite case, one should expect a soft supercritical
transition (corresponding to a bifurcation of the usual pitchfork type \cite
{bifurcation}) without a jump.

The bifurcations observed in Figs. 1(a) and, especially, in Fig.~\ref
{Fig:u(gamma)}(b) look very much like a supercritical pitchfork bifurcation,
while Figs. 1(c) and 1(d), corresponding to larger values of the parameter
$k^{2}\equiv \omega ^{2}-1$, strongly suggest that a subcritical bifurcation
(the one giving rise to a jump) takes place in these cases. In many other
runs of the simulations, not displayed here, we observed a very similar
trend, namely, to have a supercritical bifurcation for very small values of
$k^{2}$, and a transition to a subcritical bifurcation at somewhat larger
$k^{2}$. An accurate prediction of the type of the bifurcation (sub- or
super-critical) is a rather difficult issue, as it demands precise summation
of the lowest-order nonlinear corrections at the order $\avu^{2}$,
that may originate from many different terms in the above analysis.

Lastly, we note that the case shown in Fig.~\ref{Fig:u(gamma)}(a), with the
driving frequency $\omega =0.75$, which is smaller than $1$, cannot be
directly explained in the framework of the approach developed here. However,
it seems very plausible that the persistent motion of the ac-driven fluxon
in this case can be explained if one takes into regard the fact that the
kink oscillating under the action of the driving frequency $\omega <1$ emits
radiation at the second-harmonic frequency $2\omega $, provided that
$2\omega >1$ \cite{review}, which is the case in Fig.~\ref{Fig:u(gamma)}(a).
In accord with this, the effect seen in Fig.~\ref{Fig:u(gamma)}(a) is very
weak because it is accounted for by the second order of PT.

\section{Comparison of analytical and numerical results}

\label{Sec:Disc}

The main prediction of the analysis presented in the previous section is the
loss of stability of the zero-average-velocity state of the fluxon in the
ac-driven weakly damped annular uniform LJJ. As a result, the fluxon must
inevitably perform a transition to a nontrivial state with a nonzero average
velocity, which we consider as the cause for the effect revealed by the
simulations. Besides that, the analysis may, in principle, predict the
character of the bifurcation -- super- or sub-critical.

While the comparison of the particular type of the analytically predicted
bifurcation with results of the simulations is a sophisticated problem, it
is much more straightforward (and more important) to compare the
theoretically predicted and numerically found points of the transition to
the progressive motion, i.e., as a matter of fact, dependencies of the
threshold amplitude of the ac drive on various parameters of the system. A
crucially important prediction of the analysis is the fact that the
transition to the nonzero velocity occurs (within the framework of the
perturbation theory) only in those intervals of values of the ring's length
$L$ (if all the other parameters are fixed) where the condition (\ref
{threshold}) holds. Another important and, in the same time, simple
prediction is that the threshold value (\ref{threshold}) remains finite
(neither vanishes nor diverges) in the limit $\alpha \rightarrow 0$.

\begin{figure}[tb]
  \centering\includegraphics{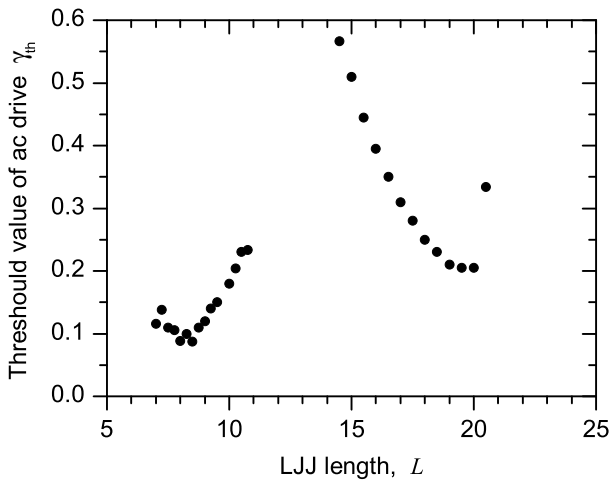}
  \caption{
    The minimum (threshold) value of the amplitude of the ac driving field, at which the progressive motion of the fluxon commences, vs. the length of the annular Josephson junction, as found from the numerical simulations. The driving frequency and dissipative constant are $\protect \omega = 1.12$ and $\protect\alpha = 0.1$.
  }
  \label{Fig:Thr(L)}
\end{figure}

Comparing the analytical and numerical results, we focused on these two
basic features. In Fig.~\ref{Fig:Thr(L)}, the threshold value of the
ac-drive's amplitude, as found from the simulations, is plotted vs. the
length $L$ of the annular junction for fixed values $\omega =1.12$ and
$\alpha =0.1$. Salient peculiarities of the dependence are existence of two
minima at
\begin{equation}
\left( L_{\min }^{(1)}\right) _{{\rm num}}\approx 8.5,\,\left( L_{\min
}^{(2)}\right) _{{\rm num}}\approx 20.5\,.  \label{min}
\end{equation}
On the other hand, the value of the wavenumber corresponding to $\omega =1.12
$ is $k\approx 0.53$ [see Eq.~(\ref{k})], and Eq.~(\ref{threshold}) then
predicts minimum values of the threshold at the points where $\sin \left(
kL\right) =-1$, i.e., at
\begin{equation}
\left( L_{\min }^{(1)}\right) _{{\rm theor}}=\frac{3\pi }{2k}\approx
8.95,\,\left( L_{\min }^{(2)}\right) _{{\rm theor}}=\frac{7\pi }{2k}\approx
20.90\,.  \label{theormin}
\end{equation}
Comparison of the numerical results (\ref{min}) with the analytical
predictions (\ref{theormin}) demonstrates very good agreement.

Figure~\ref{Fig:Thr(L)} also shows that $\gamma _{{\rm thr}}$ diverges at
some point between the two minima, and in a part of the region between them,
where the data is absent in the figure, the effect does not take place at
all (or the threshold is extremely high). This feature also agrees well with
the theoretical prediction (\ref{threshold}), which yields divergences at
points $L=\pi n/k$ with integer $n$, and nonexistence of the effect in the
intervals where the condition (\ref{condition}) does not hold. However,
unlike the minima points, detailed comparison of the numerically found and
analytically predicted positions of the divergence points is not relevant,
as the perturbation theory which was used to derive Eq.~(\ref{threshold})
clearly breaks down in a vicinity of the divergence points.

\begin{figure}[tb]
  \centering\includegraphics{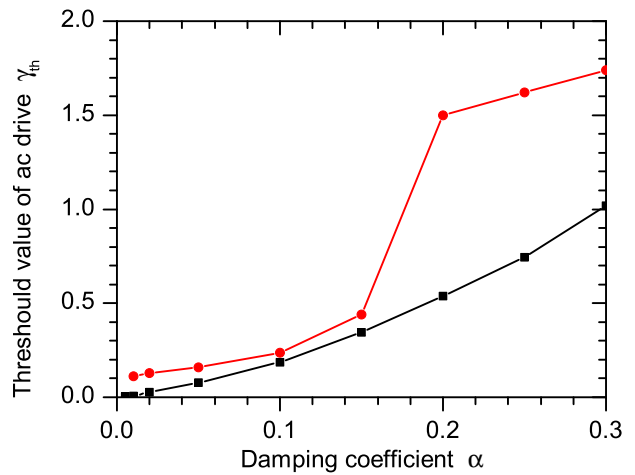}
  \caption{
    The threshold value of the amplitude of the ac driving field vs. the dissipative constant $\protect\alpha$. The length of the annular Josephson junction and driving frequency are $L=10$ and $\protect\omega = 1.12$. The upper branch of the plot corresponds to the case when simulations start with a quiescent fluxon, and the drive's amplitude is gradually increased till the fluxon starts to drift. The lower branch corresponds to the opposite case, when the simulations start with a fluxon moving under the action of a sufficiently strong drive, and the amplitude is gradually decreased till the fluxon ceases to drift.
  }
  \label{Fig:Thr(alpha)}
\end{figure}

To verify the analytical prediction for the threshold amplitude of the ac
drive as a function of the dissipative constant $\alpha $, the dependence
$\gamma _{{\rm thr}}(\alpha )$ is shown in Fig.~\ref{Fig:Thr(alpha)}. The
first noteworthy feature of the plot is strong hysteresis. For the
comparison with the analytical prediction, the relevant branch is the upper
one, which corresponds to ``sweeping up'', i.e., gradual increase of $\gamma
_{{\rm ac}}$, as precisely in this case we can detect the point at which the
zero-average-velocity state becomes unstable, and the fluxon starts its
progressive motion.

As it is evident in Fig.~\ref{Fig:Thr(alpha)}, this branch of the dependence
of $\gamma _{{\rm thr}}$ vs. $\alpha $ indeed yields a finite value in the
limit $\alpha \rightarrow 0$, as it is predicted by Eq.~(\ref{threshold});
it should be stressed, though, that accurate simulations are quite difficult
for very small $\alpha $, as relaxation of the dynamical regime to its
established form is very slow in this case. A particular value that Eq.~(\ref
{threshold}) yields for $\alpha =0$ in the case shown in Figs.~\ref
{Fig:Thr(L)} and \ref{Fig:Thr(alpha)}, i.e., $L=10$ and $\omega =1.12$, is
$\gamma _{{\rm thr}}\left( \alpha =0\right) \approx 0.2$, which roughly
agrees with what can be read off from Fig.~\ref{Fig:Thr(alpha)}.

\section{Conclusion}

\label{Sec:Conclusion}

This work presents a novel dynamical effect: systematic drift of a
topological soliton in an ac-driven weakly damped system with periodic
boundary conditions. The effect has been demonstrated in a physically
relevant model of a long Josephson junction in the form of a ring, where
fluxons play the role of kinks, and the ac drive is realized as bias current
uniformly applied to the junction. Unlike earlier considered cases of
progressive motion of the ac driven fluxon, in the present case the long
junction and the ac driving force are spatially uniform. Numerical
simulations reveal that progressive (on average) motion of the fluxon
commences if the amplitude of the ac drive exceeds a certain threshold
value. With the increase of the amplitude, both regular and chaotic
dynamical regimes are observed. The direction of the progressive motion is
randomly selected by initial conditions, and regular dynamical regimes are
characterized by strong hysteresis. The simulations demonstrate the effect
in a well-pronounced form in the case when the driving frequency exceeds,
but its rather close to, the plasma frequency of the junction.

The analytical approach to the problem was based on consideration of the
interaction between plasma waves emitted by the fluxon oscillating under the
action of the ac drive and the fluxon itself, after the waves complete the
round trip along the annular junction and hit the fluxon. In particular, a
weaker effect, which is observed in the case when the driving frequency is
smaller than the plasma frequency, may be explained by the emission of the
plasma waves at the second harmonic by the oscillating fluxon. The main
finding which the analytical consideration yields is possible instability of
the zero-average-velocity state of the fluxon interacting with its own
radiation tails. The instability sets in if the drive's amplitude exceeds an
explicitly found threshold. The analysis predicts that the effect is only
possible if the phase shift $\varphi \equiv kL$ of the radiation wave,
gained after the round trip, is such that $\sin \varphi <0$ [see Eq.~(\ref
{condition})], and the threshold amplitude strongly depends on $\varphi $,
see Eq.~(\ref{threshold}). Numerical results show a similar dependence, and
analytically predicted values of the length of the annular junction at which
the threshold has well-pronounced minima are found to be in very good
agreement with numerical findings, see Eqs. (\ref{min}) and (\ref{theormin}).
Additionally, the analysis predicts that the threshold amplitude remains
finite as the dissipative constant is vanishing, which is also confirmed by
the numerical results.

\section*{Acknowledgements}

We are indebted to M. Fistul for valuable discussions. One of the authors
(B.A.M.) appreciates financial support provided by DAAD (Deutscher
Akademischer Austaustauschdienst) under the research-visit grant No.
A/01/24657. E.G. and B.A.M. appreciate hospitality of the Physikalisches
Institut at the Universit\"{a}t Erlangen-N\"{u}rnberg.

\end{document}